\documentstyle[aps,twocolumn,header]{revtex}

\begin{document}
\draft
\title
 {\Large \bf Suggestions For Giving Talks}
\author
 {Robert Geroch}
\address
 {Enrico Fermi Institute\\
 5640 Ellis Avenue,  Chicago, Illinois 60637}
 \medskip
\date
 {September 16, 1973}
\maketitle
\bigskip

%%%%%%%%%%%%%%%%%%%%%%%%%%%%%%%%%%%%%%%%%%%%%%%%%%%%
% SECTION 1  :   CHOICE OF SUBJECT
%%%%%%%%%%%%%%%%%%%%%%%%%%%%%%%%%%%%%%%%%%%%%%%%%%%%

\section{Choice Of Subject}
As a general rule, one chooses for his subject the 
broadest and most general version that he can feel 
comfortable with. This almost always involves
broadening the scope of what one really wants to say. 
The problem is that one naturally divides the subject
into what is known and what is not known, and works 
at the interface. To your audience, however, almost 
everything is unknown: they have no feeling for where
this interface is, or how the problems at the interface 
arise from the larger context. Thus, instead of talking 
about killing tensors, one could talk about conservation 
laws for particles in general relativity 
(including perhaps, some discussion of spinning particles, 
charged particles, what happens when a particle breaks into
two particles, how this subject is related to conservation 
of stress-energy, etc.). Of course, what is the most 
general version one can feel comfortable with depends on 
how specialized one's audience is.  
(Thus, for an audience of non-relativists, perhaps 90\% 
of the discussion will be on issues one does not actively 
think about on a day-to-day basis; for an audience of 
relativists, perhaps 60\%; and for an audience of
specialists in one's own sub-field, perhaps 35\%.).

The next step is to think of a title. Most of your 
audience will probably decide whether or not to come 
based solely on this title. Ideally, one wants a
title which indicates what the subject is, what the 
level of the discussion will be, and which is lively 
and friendly without being cute. Questions and 
assertions often make good titles. Of course, 
one should use no word in the title with which one 
does not expect one's audience to be familiar. Thus, 
for an audience of relativists, ``Linearized Fields in
a Kerr Background Metric'' sounds technical, 
``Perturbations of the Kerr Solution'' sounds dull, 
and ``Black Holes are Stable'' sounds good.

%%%%%%%%%%%%%%%%%%%%%%%%%%%%%%%%%%%%%%%%%%%%%%%%%%%%%%%%%%%%%%%%%%%%
% SECTION 2 :  THE PLAN
%%%%%%%%%%%%%%%%%%%%%%%%%%%%%%%%%%%%%%%%%%%%%%%%%%%%%%%%%%%%%%%%%%%%

\section{The Plan}

Divide the various things you want to say into three or 
four messages. (Three is, perhaps, slightly better than 
four, and either is much better than any other number.). 
A message might consist, for example, of some important point
together with supporting arguments and examples, or of 
a collection of remarks which share some common
property. Each of these messages will, eventually, become 
a short talk in its own right. Assign to each message 
a title (e.g., ``The Initial-Value Formulation of General 
Relativity''), and invent, for each, a non-technical 
summary of the message in a sentence or so 
(e.g., ``The `initial time' becomes, in general relativity, a
spacelike three-dimensional surface; the `state of the the 
gravitational field at that time' becomes a pair of tensor 
fields on this surface, subject to certain constraint 
equations; the `evolution of the system' is then 
described by equations which give the change in these 
fields under changes in the spacelike surface.''). The more
cohesive each message is, the better.

It is almost always necessary, in order to obtain such an 
organization, to recast the subject into a form which is 
essentially different from the way in which you normally 
think about it. (In one's mind, the subject is innumerable
interconnected small points. In the talk, the subject will 
be three of four main points.). In particular, one often 
has to omit some details one might otherwise have wanted
to say, omit connections between certain points, or add 
material to fill out a message. The idea is that,
with three or four messages, the audience can grasp and 
hold onto the structure of the entire talk. 
(People simply are not going to come away from any talk
with more than three of four essential points.). 
The difficult thing about planning a talk, in my opinion, 
is to divide things into messages which are sufficiently 
specific and cohesive that each can be treated as a unit 
(hence, remembered by the audience), and yet sufficiently 
general that, taken together, the messages tell one's story.  
One would like to come up with several 
(hopefully, very different) organizations, and then
select the best for further refinement.

%%%%%%%%%%%%%%%%%%%%%%%%%%%%%%%%%%%%%%%%%%%%%%%%%%%%%%%%%%%%%%%%
% SECTION 3  :  The Introduction
%%%%%%%%%%%%%%%%%%%%%%%%%%%%%%%%%%%%%%%%%%%%%%%%%%%%%%%%%%%%%%%%

\section{The Introduction}

The introduction normally consists of two parts: the placing 
of one's subject into its context in the rest of physics, 
and a description of the talk itself.

Start on as general a level as is feasible. If it's 
convenient (with a general audience), you might begin with 
some remarks about a recent trend in physics as a whole. 
Or (with a more specialized audience), you might begin with 
something about the direction of recent research in a broad 
area of physics (e.g., General Relativity), or some very 
general problem toward which a substantial research effort has 
been directed. Then, very slowly, increase the specialization 
until you get to the specific subject you're going to talk 
about. There may be three or four transitions between your
starting point and the arrival at your subject. 
(Even if you feel your audience knows this material already, 
it is still worth repeating. You must fix in their
minds the broad framework into which your subject fits.). 
If you don't know of any single, natural context for your 
subject, make one up.

Throughout this discussion, emphasize the types of problems 
under attack, why they are being attacked, the methods one
uses in the attack, the reasons one thinks along these lines, 
etc. Why does one think about this subject at all?
Why is it interesting?
What has it contributed to our understanding of Nature? 
What is the present state of the subject? Where is it 
going; what can we expect in the future?
(Predictions about the future are always a good way to 
generate enthusiasm.).  It is here that one sets the 
mood for the entire talk. As clearly and forcefully as 
you can, state what the scope of your subject is, and 
set that subject in its broader context.

The next step is to reveal the plan of the talk. 
That is, one says what his three or four messages 
will be. You might give the title of each message 
(and, perhaps, write these titles on the board, 
so you can check them off as each message is delivered), 
and a few descriptive sentences on each one. 
Furthermore, one wants to tie all these messages 
together.  How do the various messages relate to each 
other, and how, taken together, do the messages 
constitute a summary of your subject? If it can 
be done conveniently, you might also reveal here what 
your general conclusions will be. In short, one gives 
a short talk on the structure of his talk.

This introduction normally consumes about one-fifth of 
the time available.

%%%%%%%%%%%%%%%%%%%%%%%%%%%%%%%%%%%%%%%%%%%%%%%%%%%%%%%%%%%%%%%%
% SECTION 4  :  The Body Of The Talk
%%%%%%%%%%%%%%%%%%%%%%%%%%%%%%%%%%%%%%%%%%%%%%%%%%%%%%%%%%%%%%%%

\section{The Body Of The Talk}

At the beginning of each message, state again what the 
title of the message is, that you're now beginning on 
that message, and, if you like, a few sentences
about the content of the message. At the end of each 
message, state that you've finished the message, and 
give your few-sentence summary of it. 
(That is, sentences which begin ``We now begin our 
discussion of...'' and ``To summarize,...''.). These
summaries, particularly, are very important: 
they give your audience a chance to think over what 
you've said, and to draw it all together. 
(They also provide an opportunity for those
who have gotten lost in that message to again pick 
up the thread of the talk.).  This transition from 
one message to the next must be abundantly clear to the
audience. One can also repeat, between certain messages, 
what the plan is, where you are currently in that plan, 
and how all the messages will tie together. 
It is crucial that, at every moment during the talk, 
your audience knows what the plan is and where
they are in that plan.

The body of each message should be a short talk in itself, 
with a clear, central objective. 
(Thus, for example, if you use in one message material from
an earlier one, you should briefly summarize that material 
first.).  A message normally consists of three to six points 
you want to make.

The mode of presentation of a message is not normally the 
way one thinks about it privately. In particular, one 
should try diligently to suppress everything which does 
not bear directly on the central objective of the message. 
Examples:
\begin{tabbing}
\= iiii) \= \kill
\> i)   \>\parbox[t]{3.1in}{By changing one's private 
notation, one can often avoid introducing most variables.} \\
\>      \> \\
\> ii)  \> \parbox[t]{3.1in}{By rearranging one's private 
definitions, one can often avoid completely the definition 
(and hence  subsequent  discussion) of certain  extraneous 
concepts.} \\
\>      \> \\
\> iii) \> \parbox[t]{3.1in}{By presenting matters from a 
different (and, perhaps, ``less correct'') point of view, 
one can often arrange things so that certain side issues 
simply do not arise.}\\
\>      \> \\
\> iv)  \> \parbox[t]{3.1in}{By introducing an analogy, 
one can often afford (i.e., without loss of clarity) to 
omit the details of a topic.  (Analogies are particularly 
valuable for topics with which there are associated
a large number of small points.) By referring to the analogy, 
the audience can answer most questions on
the topic for themselves.}\\
\>      \> \\
\> v)   \> \parbox[t]{3.1in}{In discussing a calculation, 
one need only make it it (very) clear what went in and 
what came out. (If there are several consecutive calculations, 
one should consider them as one, and just give the
input and the output of the whole mess.)}
\end{tabbing}

In short, avoid everything which leads away from your 
message. It is even sometimes necessary to say things 
which are (technically) incorrect, or which contain 
omissions, in order to accomplish this objective.

%%%%%%%%%%%%%%%%%%%%%%%%%%%%%%%%%%%%%%%%%%%%%%%%%%%%%%%%%
%  SECTION 5 :  The Conclusion
%%%%%%%%%%%%%%%%%%%%%%%%%%%%%%%%%%%%%%%%%%%%%%%%%%%%%%%%%

\section{The Conclusion}

At the end of the talk, summarize (in non-technical language) 
what your main points were, and what the context of the 
subject is. Of course, you will also tie together the messages, 
make them seem like parts of a whole. This is also a good time 
to state various questions one would like to have answered. You 
might also add, at the very end, any predictions you'd like to 
make concerning what is likely to happen in this subject in the 
future.

%%%%%%%%%%%%%%%%%%%%%%%%%%%%%%%%%%%%%%%%%%%%%%%%%%%%%%%%
%  SECTION 6 :  Visual Material
%%%%%%%%%%%%%%%%%%%%%%%%%%%%%%%%%%%%%%%%%%%%%%%%%%%%%%%%

\section{Visual Material}

Figures are easier to understand than words. Words are easier 
to understand than equations.

Say it with a figure (or graph, or table) if at all possible. 
(It is surprising how many ideas can be reduced to or 
illustrated by a figure.).  Figures should, of course, be 
simple, with all inessential details omitted. Label everything
that can be labeled, and, if the figure is on a slide, give it 
a title. You should intend that every single mark on a figure 
will be fully understood by the audience. (If they don't 
understand something, leave it off the figure.). It takes time 
to absorb a figure, so have some remarks to make about the 
figure while they're staring at it. (Even though it may be 
clear from the labeling, describe the figure and its message in
words.). Do not use several figures when one can be made to do 
the job. It's often possible to invent a single, strong figure 
(or graph, or table) which forcefully summarizes the essential 
point.

If you make a point in words, or give an argument in words, 
it's often possible to summarize it on a slide or on the 
blackboard. This might be done by writing out a sentence or 
two in full. (This endows your point with strong emphasis.).
For an argument with several steps, one might list the 
steps (one phrase for each), to bring out the structure of 
the argument. It never hurts to read aloud what is displayed.
It's not usually a good idea to try to get more than two 
sentences on a slide.
(If one is stating a theorem, for example, one might 
condense some complicated conditions into a descriptive 
phrase in quotation marks.).  The slide should, of course, 
be up far longer than it takes to read it.  One would 
normally spend at least several minutes on a slide
. (If it's to be less, perhaps the whole slide can be 
replaced by a descriptive phrase in some other slide.).

The last resort for expressing an idea is through an 
equation. In my opinion, equations should be thought 
of as tools for making a point, not as data to be
stored by the audience for their future use. 
(How many times have you actually used, in your own work, 
a detailed equation copied from a lecture?). 
Thus, an equation should be a ``picture'' which is 
presented, described in detail, discussed physically, etc.
Every symbol appearing should be defined, and, if 
necessary, discussed. All this takes time, so it is a 
good idea to set aside several minutes to treat a single 
equation. (The meaning of important symbols should be 
repeated in later equations, even though they were 
defined and discussed at their first appearance.). 
If a talk has more than five non-trivial equations 
in it, it's beginning to get equation-heavy. One can 
often simplify equations by a clever choice of 
variables (e.g., define a variable to represent 
``the effect of the gravitational field on the stress 
of the body'').  Furthermore, one can often leave 
out whole batches of terms by summarizing them with 
a phrase, e.g., `` + small terms''. 
If some terms are not going to be discussed in
detail, replace them by a word saying why they are not 
important. One can sometimes get rid of an entire equation 
by writing it symbolically, or with words replacing the 
terms. You should intend that every single mark that 
appears in an equation will be completely understood by 
your audience.

Ideally, one would like to have about ten slides 
(or blackboard presentations) in an hour talk, with 
one under discussion perhaps 70~\% of the time.

%%%%%%%%%%%%%%%%%%%%%%%%%%%%%%%%%%%%%%%%%%%%%%%%%%%%%%%%%%%%%%%%
%  SECTION 7 :  General Suggestions
%%%%%%%%%%%%%%%%%%%%%%%%%%%%%%%%%%%%%%%%%%%%%%%%%%%%%%%%%%%%%%%%

\section{General suggestions}

Speak loudly and firmly, with conviction.  
Try to make every sentence you construct a sentence 
you can be confident about.  (Don't, for example, say 
that something is sort of true, or use a weak, hesitant 
voice.  A rough physical argument is not just a sloppy 
version of some precise argument; it is a carefully 
formulated, defensible statement, beginning, for example, 
with ``In physical terms \ldots'') Use full sentences as 
you compose your thoughts or think of how to best express 
your next idea.  If you realize you've made a serious 
error, or have gotten confused, say so, and try to
straighten things out, out loud, with your audience.  
If you realize that you've said something in an unclear 
way, it's usually best to announce this fact, and that 
you'll now try to say it again.  You can usually tell 
from your audience's facial expressions when they don't 
understand, and when you're going too slowly.

Be explicit whenever you can find a way to do so.  
Avoid ``this'' and ``that'' as nouns: say 
``that tensor'' or ``that charged particle''. 
Don't say ``it'' for something that has a name.  
One can sometimes introduce an artificial explicitness .  
Thus, in a discussion, ``the ten-gram mass'' and 
``the five-gram mass'' are better labels than
``this mass'' and ``that mass'', or ``$m_1$'' and ``$m_2$''.

Try to keep the audience informed of what you're doing, 
how things fit together, where you're going, etc.  
Thus, if you give an example of an argument, say 
before you start that it is an example, and after you're 
done that it was an example and that you're now returning 
to the main discussion.

Some of your comments will be vastly more important than others.  
It is vital that you indicate this relative importance.  
The following are techniques for emphasizing a point:
\begin{tabbing}
\= iiii) \= \kill
\>   i)  \> Say the point in a single, loud, short sentence,\\
\>       \> \\
\>  ii)  \> repeat the point several times,\\
\>       \> \\
\> iii)  \> pause after making the point,\\
\>       \> \\
\>  iv)  \> say ``The following point is important:\ldots''
\end{tabbing}
By using combinations of such techniques, try to get the correct
distribution of emphasis in your presentation.

Do not allow your audience to get bored.  If they look bored, 
try to drum up some enthusiasm.  You might, for example, stop 
what you're doing and repeat, loudly, what your plan is, 
where you are in that plan, and why this problem is of interest.  
It is almost always a disaster to run over one's time.  
(The audience becomes bored and anxious to leave.  
Not only do they not learn anything after your time is up,
but they tend to lose the thread of what went before.) 
If you see that your time is up before you 've finished, 
I would suggest that you stop there, and summarize in a few
sentences.  One often prepares a few additional points 
(e.g. examples) which could be worked into the talk, 
but which are not essential.  One can decide during the 
talk whether or not to include these points, in order 
to make the time come out right.

Everyone has his own system for notes for a talk.  
I prefer a single sheet of paper with an outline 
of the talk on one side: the introduction, titles of the
messages, and conclusion are the main divisions; the five 
or so points to be made in each of these divisions 
(summarized to a phrase) are the subdivisions.  
Occasionally, one jots an equation or two on the back.

In my opinion, the most important point about answering 
questions after (during) a talk is to be completely honest.  
If someone says that an argument does not seem convincing, 
and if you have doubts about it, say that it doesn't
seem convincing to you either.  If someone catches you on 
an ``omission'', say that you omitted it to simplify the 
discussion, and fill in the missing material.  If someone 
asks something that hasn't occurred to you, say that it 
hasn't occurred to you, and whatever else you can.
If someone asks a question in the middle of the talk, 
it's usually best, after the answer, to resume by 
working your way slowly up to where you were, saying
the most distant material in a general way, getting more 
specific until you reach the point at which you were 
interrupted.  If a discussion within the audience threatens 
to take over (in the middle of the talk), one can say 
(firmly) that he would prefer to postpone discussion of 
the issue until the end.

Talking about physics does not closely resemble thinking 
about physics because the purposes in the two cases are 
entirely different.  The amount of information you emit 
is irrelevant; it's the amount you cause to be absorbed
that counts.  A talk has a clear objective, to force 
certain information into the minds of the audience.  The
idea is to direct one's entire effort to accomplish this 
objective.
Surprisingly, experience in giving talks does not, after 
the first few, seem to make much difference in one's 
ability to give a good talk.  What does make a difference, 
in my opinion, is having given serious and hard thought, 
over a period of time, to the art of speaking.  Two times 
are particularly valuable for doing this: after you have
just given a talk, and while you are listening to the talks 
of others.  (Was the material properly arranged?  What 
points were not clear?  Why?  What went over well, and 
what badly?  Did the audience understand the plan of the 
talk?  Were they bored?  At what points, and why?)

%%%%%%%%%%%%%%%%%%%%%%%%%%%%%%%%%%%%%%%%%%%%%%%%%%%%%%%%%%%%%
%  END OF THE DOCUMENT
%%%%%%%%%%%%%%%%%%%%%%%%%%%%%%%%%%%%%%%%%%%%%%%%%%%%%%%%%%%%%

\end{document}